\documentclass[twocolumn,aps]{revtex4}
\usepackage{graphicx}
\usepackage{dcolumn}
\usepackage{amsmath}
\begin{document}
\title[Short Title]{A Closed Integral Form for the Background Gauge Connection}
\author{Emmanuel T. Rodulfo and Joseph Ambrose G. Pagaran}
 %\email{cosetr@mail.dlsu.edu.ph}
 %\email{ambrose@zuma.ps.admu.edu.ph}
\affiliation{
Theoretical Physics Group,
Physics Department\\
College of Science, De La Salle University\\
2401 Taft, Manila, Philippines
}

\date{November 2000}

\begin{abstract}
By the appropriate use of the Fock-Schwinger gauge properties, we derive the closed integral form of the `point-split' non-local background gauge connection originally expressed as a finite sum. This is achieved in the limit when the finite sum becomes infinite. With this closed integral form of the connection, we obtain the same exact results in the calculation of one-loop effective Lagrangian accommodating arbitrary orders of covariant field derivatives in quantum field theory of arbitrary spacetime dimensions and of arbitrary gauge group.  Particularly, we display the one-loop effective Lagrangian for real boson fields up to 8 mass dimensions-the same result obtained when the connection was yet in the finite sum form.
\end{abstract}

\maketitle
\section{Introduction}
Methods for calculating the 1-loop effective Lagrangian based on the background field approach of Brown and Duff\cite{1} exist in the  literature.   Some of these methods extend Brown and Duff's covariantly constant field strength restriction to include higher order derivatives in the formulation\cite{2,3}. Customarily, one begins by imposing covariant restrictions on the background converting the nonlocal equation satisfied by the Green function to a quasilocal one.  An exactly soluble portion of the resulting Green function  equation is then identified and its solution is used as basis of a perturbative solution.  This can be done in the strong but slowly-varying background field approximation\cite{4,5} for then one can claim that invariants involving higher order covariant derivatives of the field strength tensor are significantly smaller than those with lower order derivatives of the same  mass-dimensionality.

It is im\/por\/tant to in\/clude higher deriva\/tive correction-\\s in the formalism since it has already been demonstrated \cite{6} that the bare Lagrangian for higher dimensional field theories should include invariants of higher dimensions in order to achieve renormalizability. Also, the effective Lagrangians of nonabelian field theories necessarily includes higher derivatives of the field. The accommodation of these has been achieved in a recent paper\cite{7} via the general form of the background gauge connection expressed as a finite sum of n covariant differentiations of the field strength tensor.
In this paper, we will review some aspects of the background field formalism, particularly the background gauge connection expressed in terms of the field strength tensor, and its $n$ covariant derivatives in finite sum form. From these, we will show that in the completely non-local limit when the finite sum becomes infinite-the background gauge connection may be expressed in a closed integral form by the appropriate use of the Fock-Schwinger gauge properties. We will obtain the same results as it was derived\cite{7} when the background gauge connection is yet in its finite sum form.

\section{The Background Field Approach}
In the background field method, the original Lagrangian, ${\cal L}(\phi)$, is expanded in $\phi$ about $A$,\cite{8}
\begin{eqnarray}\label{1}
{\cal L}(\phi+A)&=&{\cal L}(A)
 +\left.\frac{\delta {\cal L}}{\delta\phi_i}\right|_{\phi=A}\phi_i
\nonumber\\&&
\,\,\,\,\,\,\,\,\,
 +\frac{1}{2}\left.\frac{\delta^2 {\cal L}}{\delta\phi_i\delta\phi_j}\right|_{\phi=A}\phi_i\phi_j
 +\ldots
\end{eqnarray}
where $A$ is the classical background relative to which quantum fluctuations of the field $\phi$ are measured. By appealing to the classical equation of motion for the background\cite{9,10}, $\left.\frac{\delta {\cal L}}{\delta\phi_i}\right|_{\phi=A}=0$, the effective Lagrangian becomes
\begin{eqnarray}\label{2}
{\cal L}(\phi+A)={\cal L}(A)
+\frac{d}{d^Dx}\ln \int d[\phi]\,\,\mathrm{exp}\!\!\!\int d^Dx\, L
\end{eqnarray}
where
\begin{eqnarray}\label{3}
L=\frac{1}{2}\phi_i\left.\frac{\delta^2 {\cal L}}{\delta\phi_i\,\delta\phi_j}\right|_{\phi=A}\phi_j
\end{eqnarray}
is bilinear in the quantum fields that governs the one-loop effects. Eqn (\ref{2}), then, prescribes the calculation of the effective Lagrangian, $L_{\mathrm{eff}}$,
\begin{eqnarray}\label{4}
 {\cal L}_{\mathrm{eff}} = {\cal L}^{(0)}+{\cal L}^{(1)}
\end{eqnarray}
up to one-loop expansion. In comparison with (\ref{2}), the first term on the right-hand side is the classical Lagrangian with the classical background as its argument, while the second term represents the one-loop quantum correction. The effective Lagrangian (\ref{4}) serves as a generating functional of the proper functions.  It is an important tool in the renormalization process since it can lead to an ultraviolet finite field theory.

In particular, for real boson fields $\phi_i$ in $D$-dimensions, the most general form of the bilinear Lagrangian (\ref{3}) is given by \cite{1,2,3,8}
\begin{eqnarray}
    L\!=\!
    \frac{1}{2}\partial_\mu \phi^iW^{ij}_{\mu\nu}(A)\partial_\nu\phi^j
    \!+\!\phi^i_{\mu} N^{ij}_{\mu}(A)\partial_\nu\phi^j
    \!+\!\frac{1}{2}\phi^i M^{ij}\phi^j
\nonumber\\ \label{5}
\end{eqnarray}
where $W$, $N$, and $M$ are external space-time dependent source functions which, through adding total derivatives to ${\cal L}$, may be chosen to have (anti)symmetry properties:
\begin{eqnarray*}
    W^{ij}_{\mu\nu}=W^{ij}_{\nu\mu}=W^{ji}_{\mu\nu},\,\,\,
%    \label{Wij}\\
    N^{ij}_{\mu}=-N^{ji}_{\mu},\,\,\,
%    \label{Nij}\\ \label{Mij}
    M^{ij}=M^{ji}.
\end{eqnarray*}
Throughout this paper, we will be working in flat Euclidean D-dimensional spacetime metric
$\delta_{\mu\nu}$  with all indices lowered, in which
\begin{eqnarray}
    W^{ji}_{\mu\nu}=-\delta_{\mu\nu}\delta^{ij},
\,\,\,
%    \label{EWij}\\ \label{ED}
\,\,\,
    \delta_{\mu\mu}=D,
%    \\ \label{En}
\,\,\,\,\mathrm{and}\,\,\,\,
    \delta^{ii}=n.
\end{eqnarray}

The bilinear Lagrangian (\ref{5}) can be rewritten in the manifestly covariant form\cite{8}
\begin{eqnarray}\label{6}
    L=\frac{1}{2}\phi\left({\cal D}^2+X\right)\phi,
\end{eqnarray}
if one form the tensor quantities
\begin{eqnarray}
    X&\equiv& M-N_\mu N_\mu \label{X}    \\
    Y_{\mu\nu}&\equiv&
     N_{\nu,\mu}- N_{\mu,\nu}
    +[N_\mu,N_\nu]\label{Y}
\end{eqnarray}
which together with $\phi$ transforms according to
\begin{eqnarray}
    X&\longrightarrow & e^{\Lambda(x)}Xe^{-\Lambda(x)},\,\,\,\mathrm{etc}.
%    \label{Xto}\\ \label{Yto}
%    Y_{\mu\nu}&\longrightarrow &e^{\Lambda(x)}Y_{\mu\nu}e^{-\Lambda(x)}
%    \\ \phi&\longrightarrow & e^{\Lambda(x)}\phi \label{pto}
\end{eqnarray}
for some arbitrary antisymmetric matrix $\Lambda^{ij}(x)$. Here, the covariant derivative is defined by
\begin{eqnarray}
    {\cal D}_\mu \phi &\equiv&
    \partial_\mu\phi +N_\mu\phi,
\\
    {\cal D}_\mu T
      &\equiv&
    \partial_\mu T + \left[N_\mu,T\right],
\end{eqnarray}
where $T$ represents any of the tensors $Y$ and $X$, and their covariant derivatives. The background-dependent potential term X may be chosen so that it transforms covariantly.  The field strength tensor $Y_{\mu\nu}$ is defined in terms of the  background gauge connection $N_\mu(A)$  by (\ref{Y}). The one-loop effective Lagrangian is calculated from the coincidence limit of the two-point (Green) function associated with the bilinear Lagrangian (\ref{6}),\cite{11}
\begin{eqnarray}
    G(x,x')\!\!&=&\!\!\left<\phi^j(x)\phi^k(x')\right>
\nonumber\\
    \!\!&=&\!\!\frac{\int d[\phi]\,\phi^j(x)\phi^k(x')
    \exp\int d^Dx\,
    \frac{1}{2}\phi\left({\cal D}^2+X\right)\phi}
    {\int d[\phi]\,
    \exp\int d^Dx\,
    \frac{1}{2}\phi\left({\cal D}^2+X\right)\phi}.
\nonumber\\
\label{13}
\end{eqnarray}
The right-hand side is simply the second term in (\ref{2}), after making functional differentiation. The Green function (\ref{13}) satisfies the differential equation
\begin{eqnarray}
    \left[(\partial_\mu+N_\mu(x))^2+X\right]G(x,x')\!&=&\!
    \left<\phi(x)\phi(x')\right>
\nonumber\\\label{14}
    \!&=&\!-\delta(x,x').
\end{eqnarray}

Provided one can find some way of solving this nonlocal equation, the one-loop correction to the effective Lagrangian is given by\cite{11}
\begin{eqnarray}\label{15}
    {\cal L}^{(1)}
    =\frac{1}{2}\mbox{Tr}\int dX\,\,G(x,x').
\end{eqnarray}
Various schemes have been devised to address this nonlocal problem.\cite{4,5,13,14,15,16,17,18}

\section{Closed Form of the Point-Split Nonlocal Connection}
We now come to the main result of our paper. We shall solve (\ref{15}) in a similar scheme except that the background gauge connection from its finite sum form is expressed in the closed integral form.

Recently, it was shown\cite{7} that the general form of the background gauge connection is given by the finite sum
\begin{eqnarray}\label{16}
    N_\mu
    &=&e^{\Lambda(x)}\partial_\mu e^{-\Lambda(x)}
\nonumber\\&&
    +\sum^n_{q=1}
    \frac{(-1)^{1+q}}{(1+q)!}
    ({\cal D}^{q-1}Y(x))_\mu\circ (x-x')^q
\end{eqnarray}
provided that ${\cal D}^nY=0$. Here, $\Lambda(x)$  is some antisymmetric matrix. The notation used is
\vspace{-0.2cm}
\begin{widetext}%\end{widetext}
\begin{eqnarray*}
%\!\!\!\!&&
{\cal D}^{n}_{\mu_{1}\mu_{2} \ldots \mu_{n}}
\equiv
{\cal D}_{\mu_{1}}{\cal D}_{\mu_{2}} \ldots {\cal D}_{\mu_{n}}
%\,\,\,\,\mathrm{and}\\
%\!\!\!\!&&
(A \circ B)_{\mu_{1}\mu_{2} \ldots \mu_{p+q}} = \left \{
\begin{array}{ll}
A_{\mu_{1}\mu_{2} \ldots \mu_{p}} B_{\mu_{1}\mu_{2} \ldots
\mu_{p}\mu_{p+1} \ldots \mu_{q}} & \mbox{if $p \leq q$.} \\
A_{\mu_{1}\mu_{2} \ldots \mu_{q}\mu_{q+1} \ldots \mu_{p}}
B_{\mu_{1}\mu_{2} \ldots \mu_{q}} & \mbox{if $p \geq q$}.
\end{array}
\right.
\end{eqnarray*}
\end{widetext}

We shall now show that in the completely non-local limit $n\to\infty$  of (\ref{16}) may be expressed in a closed integral form appropriate for the `point-splitting procedure' needed in the calculation of the two point function $G(x,x')$  in (\ref{15}).  Once this Green function is obtained up to a certain order in derivatives, the 1-loop effective Lagrangian ${\cal L}^{(1)}$ can be determined from its coincidence limit.

If we use the gauge transformation\cite{7}
\begin{eqnarray}
\label{17}
N_\mu &\to& e^{\Lambda(x)}(\partial_\mu + N_\mu)e^{-\Lambda(x)}
\\ \label{18}
{\cal D}^q Y(x)&\to& e^{\Lambda(x)}{\cal D}^q Y(x)e^{-\Lambda(x)}
\end{eqnarray}
we are brought to the so-called Fock-Schwinger gauge (with respect to a fixed  point x') whose fundamental property is given by
\begin{eqnarray}\label{19}
    (x-x')\cdot N(x)=0.
\end{eqnarray}

Differentiating this with respect to x, we obtain
\begin{eqnarray}\label{20}
    N_{\mu}(x)+(x-x')_\mu \partial_\nu N_\mu (x)=0
\end{eqnarray}

Repeated differentiations with respect to x yields the general property
\begin{eqnarray}\label{21}
    (x-x')^{n+1}\circ{\cal D}^n N(x)=0,
\end{eqnarray}
which implies that
\begin{eqnarray}\label{22}
    (x-x')^{q}\circ{\cal D}^q Y_{\mu\nu}(x)=(x-x')^q\partial^q Y_{\mu\nu}(x).
\end{eqnarray}

By assuming that the quantities ${\cal D}^pY(x)$ and $\partial^p Y(x)$  are regular at $x'$, it can be shown\cite{12} that the previous equation (\ref{22}) implies
\begin{eqnarray}\label{23}
 (x-x')^p\circ {\cal D}^p Y_{\mu\nu}(x')=(x-x')^p \partial^p Y_{\mu\nu}(x')
\end{eqnarray}
where
\begin{eqnarray}\label{24}
 \partial^p Y_{\mu\nu}(x')
   \equiv \left[\partial^p Y_{\mu\nu}(x)\right]_{x=x'}.
\end{eqnarray}

The field strength tensor at the spacetime point $z$ is
\begin{eqnarray}
  Y_{\mu\nu}(z)=\frac{\partial N_\nu(z)}{\partial z_\mu}-\frac{\partial N_\mu(z)}{\partial z_\nu}+\left[N_\mu,N_\nu\right].\label{25}
\end{eqnarray}
Multiplying both sides by $-(z-z')_\nu$,
\begin{eqnarray}
  -(z-z')_\nu Y_{\mu\nu}(z)
&=&-(z-z')_\nu\frac{\partial N_\nu(z)}{\partial z_\mu}
\nonumber\\
&&-(z-z')_\nu\left[N_\mu,N_\nu\right]
\nonumber\\&&
+(z-z')_\nu\frac{\partial N_\mu(z)}{\partial z_\nu}
\label{26}
\end{eqnarray}
and applying the properties (\ref{19}) and (\ref{20}) of the Fock-Schwinger gauge to the second and first terms of the right-hand side, respectively, we obtain
\begin{eqnarray}
\!\!\!-(z\!-\!z')_\nu Y_{\mu\nu}(z)\!=\! N_\mu(z)
\!+\! (z\!-\!z')_\nu\frac{\partial N_\mu (z)}{\partial (z\!-\!z')_\nu}.
\label{27}
\end{eqnarray}

To obtain the desired closed integral form of $N_\mu$, we rescale the manifold through
\begin{eqnarray}
 z=\alpha x. \label{28}
\end{eqnarray}

Equation (\ref{27}) becomes
\begin{eqnarray}
&& N_\mu(\alpha x) + \alpha(x-x')_\nu\frac{\partial N_\mu (\alpha x)}{\partial [\alpha(x-x')_\nu]}
\nonumber\\&&
\,\,\,\,\,\,\,\,\,\,\,\,\,\,\,\,
=-\alpha(x-x')_\nu Y_{\mu\nu}(\alpha x).
\label{29}
\end{eqnarray}

It is implied that in the `point-splitting' procedure, $f(x)$ is equivalent to $f(x-x')$ for  some arbitrary fixed reference point $x'$. The left-hand side of (\ref{29}) can be written as a total differential with respect to $\alpha$:
\begin{eqnarray}
&& \frac{d}{d\alpha}\left[\alpha N_\mu(\alpha x)\right]
\nonumber\\ &&
\,\,\,\,\,\,\,\,\,\,\,\,\,\,\,\,
=N_\mu(\alpha x)
 + \alpha\frac{\partial N_\mu (\alpha x)}{\partial [\alpha x_\nu]}
\frac{\partial}{\partial \alpha}\left(\alpha x_\nu\right)
\label{30}
\end{eqnarray}
where the point $x$ is `point-splitted' first into $x-x'$ before applying partial differentiation  by chain rule to recover the left-hand side of (\ref{29}). Upon integrating (\ref{29}) and (\ref{30}) over $\alpha$,
\begin{eqnarray}
\int^1_0 d\alpha\,\, \frac{d}{d\alpha}\!\left[\alpha N_\mu (\alpha x)\right]
 = -\int^1_0 d\alpha \,\,\alpha(x-x')_\nu Y_{\mu\nu}(\alpha x).
\nonumber\\
\label{31}
\end{eqnarray}
We find the closed form for the connection
\begin{eqnarray}\label{32}
    N_\mu(x)=\int^1_0 d\alpha \,\alpha Y_{\mu\nu}(\alpha x)(x-x')
\end{eqnarray}
\vspace{-0.1cm}
which already anticipates the point-splitting\cite{1} procedure to the spacetime points $x$ and $x'$.
This can be expanded to any order since the $Y_{\nu\mu}(\alpha x)$ can be  expanded in a Taylor series
\begin{eqnarray}
    Y_{\nu\mu}(x')
    \!=\!e^{(x\!-\!x')\cdot{\cal D}_x}
    \left[Y_{\nu\mu}(x)\right]_{x=x'}
    \equiv
    e^{(x\!-\!x')\cdot{\cal D}_x}Y_{\nu\mu}(x').
\nonumber\\
\label{33}
\end{eqnarray}

This is simply an infinite series in powers of $x-x'$.  Thus, a rescaling of the manifold $x\to\alpha x$ will not affect the coefficients of expansion.  We can then write $Y_{\nu\mu}$ as
\begin{eqnarray}
    Y_{\nu\mu}(\alpha x)
    \!=\!e^{\alpha(x\!-\!x')\cdot{\cal D}_{\alpha x}}
    \left[Y_{\nu\mu}(x')\right]
    \!=\!e^{\alpha(x\!-\!x')\cdot\partial_{\alpha x}}
    \left[Y_{\nu\mu}(x')\right]
\nonumber\\\label{34}
\end{eqnarray}
where in the last step, we have used the Fock-Schwinger property. Eqn (\ref{32}) and (\ref{34})  is an alternative expression for $N_\mu(x)$ in (16), which can be evaluated to any desired order of the derivatives.Formally, this alternative nonlocal expression for $N_\mu(x)$ may be substituted into the Green function equation (\ref{13}) that satisfies (\ref{14}), in which (\ref{15}) may then be solved using various methods.\cite{3,4,5,13,14,15,16,17,18} For instance, if one sets the first covariant derivatives of the field  strength tensor while retaining the first covariant derivative of the matrix potential,
\begin{eqnarray}\label{35}
  {\cal D}_\rho Y_{\mu\nu} = 0 \,\,\,\, \mathrm{and} \,\,\,\, {\cal D}_\rho X=0
\end{eqnarray}
then, if one follows the momentum space techniques of Refs \cite{1} and \cite{3}, Eqn (\ref{15}) becomes
\begin{eqnarray}
    {\cal L}^{(1)}\!\!=\!\!\frac{\hbar}{2(2\pi)^D}
 \mathrm{Tr}\!
 \int\!\!dX\! \int\!\!d^Dp\,
G_0(p)\sum^\infty_{q=0}(\Delta_1(p)\,\,G_0(p))^q,
\nonumber\\
\label{36}
\end{eqnarray}
where\cite{1,3}
\begin{widetext}
\begin{eqnarray}
    G_0(p)
    \!\!\!&=&\!\!\!\int^\infty_0 ds\,\,\,
    \!\exp\!\left\{
    Xs
    +\frac{1}{2}\mbox{tr}\ln\sec(iYs)
    +({\cal D}X)\!\cdot\!(iY)^{-3}
    \left[\tan(iYs)-iYs\right]\!\cdot\!({\cal D}X)
    \right.
    \nonumber\\&&
    \left.
    +2i({\cal D}X)\!\cdot\! Y^{-2}\left[1-\sec (iYs)\right]\!\cdot\! p
    +\frac{1}{2}p\!\cdot\! Y^{-2}2iY^{-1}\tan(iYs)\!\cdot\! p
    \right\}.
\label{37}
\end{eqnarray}
\end{widetext}

The leading term $(q=0)$  of (\ref{36}) is
\begin{eqnarray}\label{38}
    \Delta_0(p)G_0(p)=-1,
\end{eqnarray}
which is analogous to (\ref{14}). Here,
\begin{eqnarray}
    \Delta_0(p)
    &=&-p^2+X-Y_{\mu\nu}p_\mu\frac{\partial}{\partial p_\nu}
\nonumber\\ &&
    -i({\cal D}_{\mu}X)\frac{\partial}{\partial p_\mu}
    +\frac{1}{4}Y^2_{\mu\nu}\frac{\partial^2}{\partial p_\mu\partial
    p_\nu}.\label{39}
\end{eqnarray}
To recover the free Euclidean propagator
\begin{eqnarray}
    \lim_{A\to 0} G_0(p)
    %=\int^\infty_0 ds\,\,\,e^{-(p^2+m^2)s}
     =\frac{1}{p^2+m^2}.
\end{eqnarray}
one simply switches the background off , i.e. $N_\mu\to 0$, $X\to -m^2$.

The succeeding terms when $q>0$ in (\ref{36}) can be calculated from the exact result (\ref{37}). This then provides the unrestricted Green function $G$ that accommodates all covariant derivative corrections through the perturbative expansion
\begin{eqnarray}
    G(p)&=&-(\Delta_0(p)
      +\Delta_1(p))^{-1}
\nonumber\\ \label{41}
        &=&G_0(p)\sum^\infty_{q=0}(\Delta_1(p) G_0(p))^q,
\end{eqnarray}
where $\Delta_1(p)$ is the nonsoluble part of the operator\cite{7}
\begin{widetext}%\end{widetext}
\begin{eqnarray}
    %&&\!\!\!\!\!\!\!\!\!\!\!
\Delta_1(p)&=&
    \left.
     \sum^{l-1}_{q=2}\frac{(-i)^q}{q!} {\cal D}^qX\!\circ\!
    \left(\!\frac{\partial}{\partial p}\!\right)^q
    \!+\!\sum^{n-1}_{q=1}\frac{3(-i)^{2+q}(1+q)}{(2+q)!}
        \left[\!
        {\cal D}^{(q)}\cdot Y
        \!\right]\circ\left(\!\frac{\partial}{\partial p}\!\right)^q
    \right.
    \nonumber\\&&\;
   \!+\!\sum^{n-1}_{q=1}\frac{2(-i)^{q}(1+q)}{(2+q)!} p\cdot
        \left[\!
        {\cal D}^{(q)}\circ 
                      \left(\!\frac{\partial}{\partial p}\!\right)^q
        \!\right]
    \nonumber\\
    &&\;
    +
    \sum^{n-1}_{r=1}\sum^{n-1}_{q=1} \left(1-\delta_{r_0}\delta_{q_0}\right)
       \frac{(-i)^{2+r+q}(1+r)(1+q)}{(2+r)!(2+q)!}
        \left[\!
    {\cal D}^{r}Y\!
        \circ\left(\!\frac{\partial}{\partial p}\!\right)^{1+r}\!
    \right]\!\cdot\!
    \left[\!
    {\cal D}^{q}Y(x')
        \!\circ\!\left(\!\frac{\partial}{\partial p}\!\right)^{1+q}\!
        \right]
%    \nonumber\\
%    &&\;
%    \left.
%    +\sum^{n-1}_{q=0}\frac{2(-i)^q(1+q)}{(2+q)!}p\cdot\!
%       \left[
%    {\cal D}^{q}Y(x')\circ\left(\frac{\partial}{\partial p}\right)^{1+q}
%       \right]
%    \right.
\label{42}
\end{eqnarray}
\end{widetext}
appearing in the (\ref{36}), i.e.,$\Delta(p)=\Delta_0(p)+\Delta_1(p)$. For strong and slowly varying background fields, $\Delta_1\ll \Delta_0$  the expansion (\ref{35}) is assured to be convergent. The 1-loop effective Lagrangian (\ref{36}) is then calculated from the coincidence limit of the 2-point function in coordinate space. Corresponding to the leading term $(q=0)$ in (\ref{36}) with (\ref{37}) may be shown to be
\begin{widetext}%\end{widetext}
\begin{eqnarray}
  {\cal L}^{(1)}_0
=\pm\frac{\hbar}{2(4\pi)^{D/2}}
\mathrm{Tr}
 \int^\infty_0 ds\,s^{-1-D/2}
e^{-m^2s}
%\nonumber\\&&\times
 \left[
e^{+s{\cal X}} %\right.
e^{-\frac{1}{2}\mathrm{tr}\ln (iYs)^{-1}\sin iYs}
%\nonumber\\ &&\,\,\,\,\,\,\,\times
%\left.
e^{+\dot{{\cal X}}\cdot (iY)^{-3} \left(2\tan \frac{iYs}{2} - iYs\right)\cdot\dot{{\cal X}}}
%
%e^{+\dot{{\cal X}}\cdot (iY)^{-3} (iYs) \cdot\dot{{\cal X}}}
%
 \right]
-e^{{\cal X}_0s}
%\nonumber\\
\label{43}
\end{eqnarray}
\end{widetext}
where $X$ in (\ref{37}) is replaced with $X= -m^2 + {\cal X}$ and $X_0$ represents the zero reference of the background potential $X$.   Note that the covariant derivatives of $Y_{\mu\nu}$ is absent from this result.  The 1-loop effective Lagrangian derived by Brown and Duff\cite{1} and the Lagrangian of Schwinger\cite{19} come as special cases of (\ref{43}).   Brown and Duff's result, for instance, does not involve ${\cal D}X$ while that of Schwinger is good only for QED.  Because of our closed expression (\ref{32}) to (\ref{34}) for $N_\mu(x)$ the one-loop effective Lagrangian (\ref{43}) can now be evaluated to any desired order of the derivatives but the ensuing proper time $(s)$ and momentum $(p)$ integrations are expected to become more and more tedious. Here we simply display the one-loop correction to the effective action up to eight mass dimensions that can be extracted from (\ref{43}),
\begin{widetext}
\begin{eqnarray}
    {\cal L}^{(1)}
    &=& \frac{\hbar}{2(4\pi)^{D/2}}
    \mbox{Tr}
    \left\{
    \frac{\Gamma(1-D/2)}{m^{2-D}}\left({\cal X}-{\cal X}_0\right)
    \right.\nonumber \\ \nonumber&&
    +\frac{\Gamma(2-D/2)}{m^{4-D}}
    \left[\frac{1}{2}({\cal X}^2-{\cal X}_0^2)+\frac{1}{12}Y_{\mu\nu}Y_{\mu\nu}\right]
    \\ \nonumber&&
    +\frac{\Gamma(3-D/2)}{m^{6-D}}
    \left[\frac{1}{6}({\cal X}^3-{\cal X}_0^3)
    -\frac{1}{12}{\cal X}_{.\mu}{\cal X}_{.\mu}
    +\frac{1}{12}{\cal X} Y_{\mu\nu}Y_{\mu\nu}
%    \right.\\ \nonumber&&\left.
%    \;\;\;\;\;\;\;\;\;\;\;\;\;\;\;\;\;\;\;\;\;\;\;\;\;
    -\frac{1}{90}Y_{\mu\nu.\nu}Y_{\mu\rho.\rho}
    -\frac{1}{360}Y_{\mu\nu.\rho}Y_{\mu\nu.\rho}
    \right]
    \\ \nonumber&&
    +\frac{\Gamma(4-D/2)}{m^{8-D}}
    \left[\frac{1}{24}({\cal X}^4-{\cal X}_0^4)
    +\frac{1}{24}{\cal X}^2Y_{\mu\nu}Y_{\mu\nu}
    \right.
%   \\ \nonumber&&
%    \;\;\;\;\;\;\;\;\;\;\;\;\;\;\;\;\;\;\;\;\;\;\;\;\;
    +\frac{1}{288}Y_{\mu\nu}Y_{\mu\nu}Y_{\rho\sigma}Y_{\rho\sigma}
%\\&&
    \left.\left.
%    \;\;\;\;\;\;\;\;\;\;\;\;\;\;\;\;\;\;\;\;\;\;\;\;
    +\frac{1}{360}Y_{\mu\nu}Y_{\nu\rho}Y_{\rho\sigma}Y_{\sigma\mu}
    +\ldots\right]+\ldots
    \right\}.
\nonumber\\
\end{eqnarray}
\end{widetext}
Our formulation is quite general in that it is done in arbitrary number of covariant field derivatives, in arbitrary spacetime dimensions, and in arbitrary gauge, but still confined to the one-loop approximation. We will generalize further, in a future paper, by extending this one-loop effective Lagrangian calculation to arbitrary number of loops.

\end{document}